\documentclass[prl, twocolumn,showpacs]{revtex4}
\usepackage{graphicx}

\begin{document}
  \title{Probing Kitaev Models on Small Lattices}
  \author{Han-Dong Chen}
  \author{B. Wang}
  \author{S. Das Sarma}
  \affiliation{Condensed Matter Theory Center and Center for Nanophysics and Advanced Materials, Department of Physics, University of Maryland, College Park, MD 20742-4111}
\begin{abstract}
We address the following important question: how to distinguish
Kitaev models experimentally realized on small lattices from other
non-topological interacting spin models. Based on symmetry
arguments and exact diagonalization, we show that a particularly
characteristic pattern of spin-spin correlations survives despite
finite size, open boundary and thermal effects. The pattern is
robust against small residual perturbing interactions and can be
utilized to distinguish the Kitaev interactions from other
interactions such as antiferromagnetic Heisenberg interactions.
The effect of external magnetic field is also considered and found
to be not critical.
\end{abstract}
\pacs {05.30.Pr, 37.10.Jk, 03.67.Lx}
  \maketitle

A great deal of interest \cite{1,2,3} has recently focused on the
possible realization of exotic anyonic quasi-particle statistics
in two-dimensional interacting topological systems. Much of this
interest arises from the intrinsic fundamental significance of
anyons, which are neither fermions nor bosons, and are thus
theoretically allowed only in two dimensions where particle
exchange is characterized by the braid group rather than the
permutation group (as in ordinary three dimensional systems). The
possibility of carrying out fault-tolerant topological quantum
computation \cite{1,2,3,4} using anyonic braiding is another key
reason for the current interest in the subject.

Broadly speaking, there are two alternative and complementary
routes which have been pursued in the literature for the physical
realization of the topological phase and anyonic quasi-particles.
One route \cite{1} is studying physically occurring quantum states
in nature which are believed (or perhaps conjectured) to be
anyonic in character because their low-energy properties are
thought to be well-described by some model topological quantum
field theory. The prime example of such a situation is the 5/2
fractional quantum Hall state \cite{5} which is widely considered
to belong to the $(SU_2)_2$ conformal field theory \cite{1}. A
great deal of experimental \cite{6} and theoretical \cite{7} work
is currently being pursued all over the world with the goal of
realizing the fractional quantum Hall topological qubit using the
non-Abelian anyonic quasi-particle braiding statistics \cite{1}.
Closely related to the 5/2 topological fractional quantum Hall
state is the chiral p-wave superconducting state \cite{8} in
$SrRuO_3$ or cold atoms where anyonic Majorana particles may
exist. The second route to the realization of the topological
phase, pioneered by Kitaev \cite{2,3} and the subject matter of
our work, involves the explicit construction of model spin
Hamiltonians which, by design, have topological ground states with
Abelian or non-Abelian anyonic quasi-particle excitations. In
addition to the Kitaev model, topological matter in this category
of model Hamiltonian systems includes the Levin-Wen model
\cite{9}. We note the interesting (and somewhat ironic) dichotomy
between the two classes of topological matter discussed above: in
the first category, the physical systems (e.g. the 5/2 quantum
Hall state) exist in nature, but may not be topological, whereas
in the second category the model Hamiltonians are, by design,
topological, but may not exist in nature!

In this letter, we consider the important issue of the extent to
which the topological character of the Kitaev model can be
preserved in a finite size system (e.g. a few plaquettes only),
which could possibly be physically implemented in an atomic system
such as an ion trap lattice with 20-30 ions or a cold atom (or
molecular) optical lattice with suitable interactions. We do not
discuss the logistical question of how to construct such a
lattice, which has much been discussed in the recent literature
\cite{10}. Our focus here is on the deep and fundamental question
of which characteristic properties of the thermodynamic Kitaev
model could be manifested in a finite size lattice of only a few
plaquettes. We find, rather surprisingly, that a few plaquettes
may be enough to preserve several characteristic features of the
Kitaev model. An important possible application of our results
could be the development of techniques to check whether a
particular finite size atomic (or ionic or molecular) system is
likely to manifest topological behavior. Given the great recent
success of atomic systems as emulators of well-known strongly
correlated model Hamiltonians (e.g. the Bose-Hubbard model and the
fermionic Hubbard model), it seems likely that a small finite size
Kitaev model made of ion traps or polar molecules could lead to
the emulation of a topological phase in the laboratory. Our
theoretical results, establishing the impressive robustness of
topological matter, arising from the large number of non-trivial
independent conserved operators in the model and quantitatively
verified by explicit exact diagonalizaiton calculations, apply to
both the Kitaev honeycomb lattice and the toric code. In addition
to the finite size behavior of the Kitaev model, we also study the
robustness of such small systems to possible perturbing
interactions and external magnetic fields, establishing
quantitative criteria for the observation of the characteristic
thermodynamic Kitaev model features in realistic small atomic
systems.

In this work, we focus on the measurements of local objects such
as spin-spin correlations and magnetization in an open boundary
system. We are motivated by the existence of a large set of local
conserved quantities in the Kitaev models \cite{2,3}. Based on
symmetry arguments, we are able to conclude that the local
conserved quantities impose very strict constraints on spin-spin
correlations \cite{Chen2008}, and an extremely characteristic
pattern emerges in the spatial distribution of spin-spin
correlations. More interestingly, this pattern is protected
against small size, open boundary, and thermal effects. It is also
robust against small perturbing interactions that may be present
in realistic experimental setups. Our main results are summarized
in Fig.\ref{FIG-kitaev} where the characteristic ordered emergent
correlation pattern of the Kitaev model are compared with the
messy results of the anisotropic Heisenberg model shown in
Fig.\ref{FIG-Heisenberg}.

We first study the Kitaev model \cite{3} on a honeycomb lattice
sketched in the top left panel of Fig.\ref{FIG-kitaev},
\begin{equation}
   H =  \sum_{\alpha=x,y,z}\quad\sum_{\alpha-bonds} J_\alpha \sigma^\alpha_b \sigma^\alpha_w, \label{EQ-Kitaev-H}
\end{equation}
where the subscripts $b$ and $w$ denote the two end sites(black or
white) of nearest-neighbor bonds and $\sigma$'s are the Pauli
matrices. This model has two phases \cite{3}. The gapped phase has
Abelian anyons as excitations, whereas the gapless one supports
non-Abelian anyonic excitations in the presence of an external
magnetic field. In this work, we study the case of $J_x=0.4,
J_y=0.4$ and $J_z=1.0$ which is gapped. Our symmetry argument
holds for both gapped and gapless phases.

For each plaquette, there is one conserved quantity. For instance,
for the plaquette enclosed by sites $1$-$6$, the operator
$W_p=\sigma^y_1 \sigma^z_2 \sigma^x_3 \sigma^y_4 \sigma^z_5
\sigma^x_6$ is conserved \cite{3}. These conserved quantities have
profound implications for the physics of the Kitaev model
\cite{3,Baskaran2007,Feng2007, Chen2008}. For spin-spin
correlations, it is always possible to find a conserved quantity
that flips one spin without changing the others, unless the
following two conditions are both satisfied \cite{Chen2008}:
\begin{itemize}
    \item The two spins are nearest neighbors;
    \item Their components agree with the bond direction.
\end{itemize}
The spin-spin interaction terms in Eq.(\ref{EQ-Kitaev-H}) satisfy
the above two conditions. If the conservation law applies, the
correlation functions vanish identically unless the above
conditions are both satisfied.

\begin{figure}
  \includegraphics[width=2.5in]{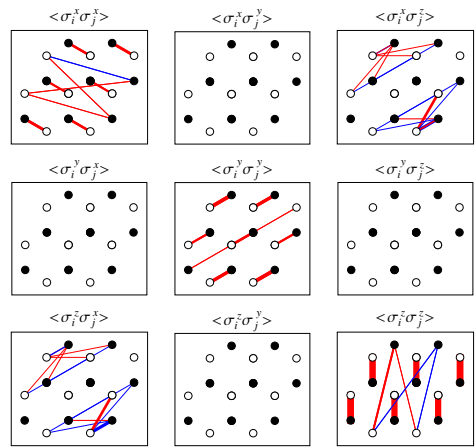}
  \caption{Spin-spin correlation functions of the 16-site Kitaev model in a pure ground state. The coupling strengths are $J_x=0.3, J_y=0.4$ and $J_z=1.0$. Red(Blue) bond denotes negative(positive) correlation. Bond thickness is proportional to the magnitude of the correlation. Empty bonds denote zero correlations.}  \label{FIG-pure-correlation}
\end{figure}

However, for the open boundary case, the boundary terms, such as
$W=\sigma^z_1\sigma^y_2$ in the 16-site lattice of
Fig.\ref{FIG-kitaev}, may not commute with each other. For
instance, $[\sigma^z_1\sigma^y_2, H]= [\sigma^x_2\sigma^z_3
\sigma^z_7, H]=0$ but $[\sigma^z_1\sigma^y_2,
\sigma^x_2\sigma^z_3\sigma^z_7]\neq 0$. Therefore, in a pure
ground state, some of the symmetries involving the boundary spins
might be broken, and consequently the spin-spin correlation
functions involving the boundary spins can have finite values. In
the 16-site lattice of Fig.\ref{FIG-kitaev}, only the $4$-th and
$10$-th sites are not on the boundary and the remaining $14$ sites
are all boundary sites. In Fig.\ref{FIG-pure-correlation}, we plot
the correlation functions in a typical pure ground state of the
$16$-site Kitaev model with parameters $J_x=0.3$, $J_y=0.4$ and
$J_z=1.0$. As expected, finite correlations are found between the
boundary spins. Furthermore, the ground state is 16-fold
degenerate. This can be understood based on the exact mapping
introduced in Ref. \cite{Chen2007,Chen2008}. Sites $2$, $7$, $12$,
$16$ have dangling Majorana fermions, each of which contributes a
factor of $\sqrt{2}$ to the ground state degeneracy. Also, each
horizontal row of the $z$-bonds has a $Z_2$ degree of freedom.
Combining all these contributions, we obtain the degeneracy
$(\sqrt{2})^4\times 2^2=16$.

\begin{figure}
\includegraphics[width=3in]{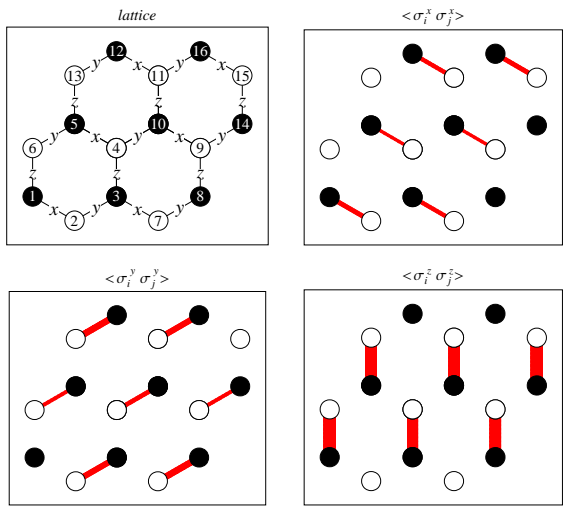}
\caption{Top-left: Honeycomb lattice of $16$ sites and three types
of bonds. Others: spin-spin correlations in the low temperature
thermal equilibrium state. All other components such as $\langle
\sigma^x \sigma^y\rangle$ vanish identically.}\label{FIG-kitaev}
\end{figure}
\begin{figure}
\includegraphics[width=3.4in]{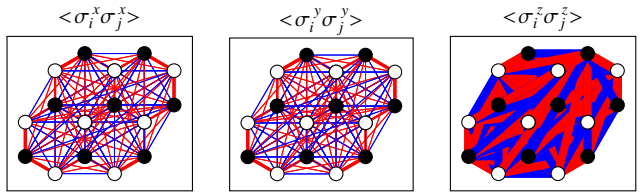}
\caption{Spin-spin correlations for the anisotropic Heisenberg
model $H= \sum_{<bw>} \sum_\alpha J_\alpha
\sigma^\alpha_b\sigma^\alpha_w$ with parameters $J_x=0.3, J_y=0.4$
and $J_z=1.0$. This is dramatically different from the case of
Fig.\ref{FIG-kitaev} of Kitaev model. }\label{FIG-Heisenberg}
\end{figure}

Since the ground state is degenerate, different pure ground state
wavefunctions lead to different spin-spin correlation functions.
Therefore, it is important to control the experimental realization
of the ground state. One interesting and simple situation is the
thermal equilibrium state instead of a pure state. For a thermal
equilibrium state
at zero temperature, the density matrix is $\rho \propto
\sum_{|g.s.\rangle} |g.s.\rangle\langle g.s.|$, where the
summation is over all degenerate ground states $\{|g.s.\rangle\}$.
In this symmetric mixed state, the broken symmetries are restored,
and one would expect correlation functions to vanish unless the
two conditions are satisfied. This can be easily seen from the
exact diagonalization results plotted in Fig.\ref{FIG-kitaev}. We
thus obtain our main result. The spin-spin correlation functions
of the Kitaev model on the honeycomb lattice are extremely short
ranged and anisotropic. As a comparison, we plot the correlation
functions of the anisotropic Heisenberg model on the same
$16$-site lattice in Fig.\ref{FIG-Heisenberg}. In this case, the
correlation functions are all over the real space and dramatically
different from the case of Kitaev model in Fig.\ref{FIG-kitaev}.

\begin{figure}
\includegraphics[width=3.4in]{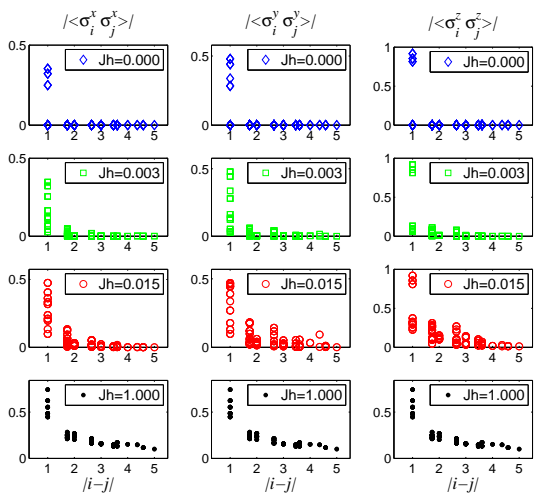}
\caption{Spin-spin correlation functions on a lattice of $16$
sites. Top three panels are pure Kitaev model and the Bottom three
panels are pure antiferromagnetic Heisenberg model. Middle six
panels are Kitaev model with residue Heisenberg interactions. The
parameters for Kitaev model are $J_x=0.3, J_y=0.4,
J_z=1.0$.}\label{FIG-K-H}
\end{figure}

Because unwanted perturbing interactions are inevitable in any
experimental realization, it is necessary to study their effects.
In particular, we consider the uniform antiferromagnetic
Heisenberg interaction $H_{res}=J_h  \sum_{<bw>}
\sum_\alpha\sigma^\alpha_b \sigma^\alpha_w$. Although this
perturbation destroys the local conserved quantities, the pattern
in Fig.\ref{FIG-kitaev} survives when the residual interaction is
a few percent of the coupling strength $J_x, J_y,J_z$.  In
Fig.\ref{FIG-K-H}, we plot the calculated correlations in the
presence of antiferromagnetic perturbation as functions of the
distance between two spins. In the top panels, we plot the results
of a pure Kitaev model. One can only find finite correlations for
some of the nearest neighboring bonds, as we discussed previously.
As we increase the perturbation to $J_h=0.003$, which is $1\%$ of
$J_x$, small correlations start to develop between
next-nearest-neighbors and next-next-nearest neighbors.
Nevertheless, the dimerization along $z$-bonds is still very
strong, {\it i.e.}, the difference between strong and weak
$\langle\sigma^z \sigma^z\rangle$ correlations remains evident. As
the perturbation further increases to $5\%$ of $J_x$
($J_h=0.015$), more long range correlations emerge and reach as
high as about $30\%$ of the strongest correlations of the
nearest-neighbor bonds. However, it still has a much shorter tail
than the pure Heisenberg model, which is shown in the bottom
panels. Furthermore, the difference between strong and weak
$\langle\sigma^z \sigma^z\rangle$ correlations is still visible.
Therefore, we conclude that it is necessary to control any
residual interactions within a few percent of the Kitaev coupling
strength to successfully observe the characteristic Kitaev pattern
depicted in Fig.\ref{FIG-kitaev}.

\begin{figure}
  \includegraphics[width=2.5in]{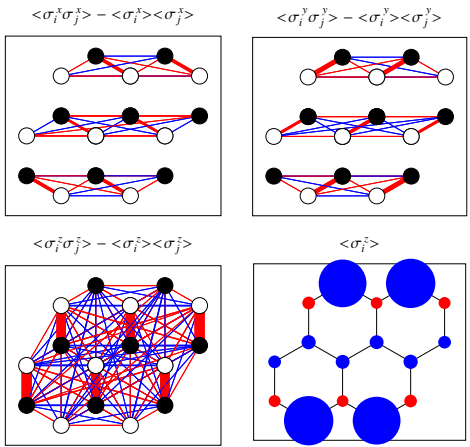}
  \caption{Spin-spin correlations and spin moment in the low temperature thermal equilibrium state when a uniform magnetic field $B_z=0.1$ along the $z$ direction is applied. In the right-bottom panel, red(blue) denotes negative(positive) moment. The size of dot denotes the magnitude of spin moment. $\langle \sigma^z_2\rangle =0.78$.\label{FIG-kitaev-B}}
\end{figure}

\begin{figure}
   \includegraphics[width=2.5in]{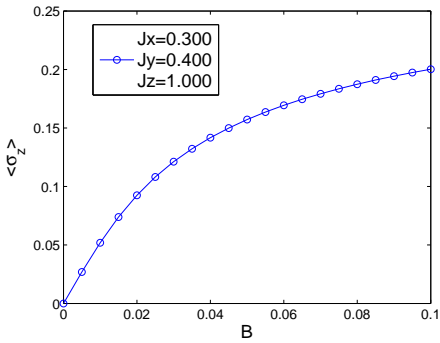}
    \caption{Magnetization $\sum_i\langle \sigma^z_i\rangle/16$ in Kitaev model as a function of uniform external magnetic field $B_z$ along the $z$ direction.}\label{FIG-kitaev-Sz-B}
\end{figure}

We now turn to another important effect, namely the effect of an
external magnetic field. When an external magnetic field is
applied, the conserved quantities defined on plaquettes are no
longer good quantum numbers. However, other conserved quantities
defined on the zig-zag chains might survive. When the field is
along the $z$-direction, the products of $\sigma^z$ on the
horizontal zig-zag chains, {\it e.g.}
$\sigma^z_1\sigma^z_2\sigma^z_3\sigma^z_7\sigma^z_8$, still
commute with the full Hamiltonian and with each other.
Consequently, the correlations between two spin components along
$x$ or $y$ directions can have finite values only if they belong
to the same horizontal zig-zag chains, as seen in the first two
panels of Fig.\ref{FIG-kitaev-B}. Longer range correlations are
developed in the $z$-components. As long as $B$ is small compared
with $J_z$, $z$-bonds are still dominated by singlets formed
between two end spins. Overall, the characteristic pattern of
Fig.\ref{FIG-kitaev} is clearly visible in Fig.5. On sites
$2,7,12,16$, where dangling Majorana fermions exist when $B=0$,
sizeable spin moment is induced along the field direction, as
plotted in the third panel of Fig.\ref{FIG-kitaev-B}. Significant
magnetization along the field direction is thus observed even for
a small magnetic field, as shown in Fig.\ref{FIG-kitaev-Sz-B}.
This is opposite to the case of anisotropic Heisenberg model,
where a spin gap prevents the magnetization of spins at low
temperature.

At finite temperature, excited states will also contribute to the
correlation functions. Fortunately, the symmetry argument holds
not only for the ground state but also for excited states. The
pattern of Fig.\ref{FIG-kitaev} is thus protected by the local
symmetries, and thermal fluctuations have no effect on it.

\begin{figure}
   \includegraphics[width=3.4in]{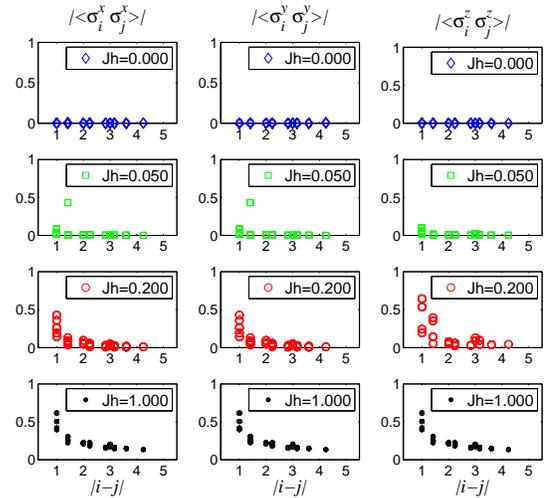}
    \caption{Correlations in toric code model of Eq.(\ref{EQ-toric}). Top three panels are pure toric code model. The second and third rows are toric mode with uniform antiferromagnetic Heisenberg interactions. The last row is pure Heisenberg model on square lattice. The coupling strength of toric code is $J=1$.}\label{FIG-toric}
\end{figure}

Finally, we also study an equivalence of the Kitaev toric code
\cite{Wen2003,Nussinov2007}. The model is defined on a square
lattice,
\begin{equation}
  H_{toric} = \sum_{\vec{r}} J \sigma^x_{\vec{r}}\sigma^y_{\vec{r}+\hat{e}_x}
  \sigma^x_{\vec{r}+\hat{e}_x+\hat{e}_y}\sigma^y_{\vec{r}+\hat{e}_y}\label{EQ-toric}
\end{equation}
where $\vec{r}$ is the lattice point of square lattice spanned by
$\hat{e}_x$ and $\hat{e}_y$. This model proposed by
Wen\cite{Wen2003} was shown to be equivalent\cite{Nussinov2007} to
the toric code of Kitaev\cite{2}. The terms in Eq.(\ref{EQ-toric})
commute with each other and form a large set of local conserved
quantities. It is thus possible to apply similar symmetry
arguments and obtain similar constraints on spin-spin correlation
functions. However, in this model, the symmetry argument does not
apply to some bonds near the four corners of the square lattice.
Nevertheless, as we can see in the first row of
Fig.\ref{FIG-toric}, spin-spin correlations vanish or are
negligibly small. As a perturbing Heisenberg interaction is
introduced, small correlations start to emerge. When $J_h=0.2 J$,
the spin-spin correlations are already dominated by the perturbing
interactions, as shown in the third and fourth rows in
Fig.\ref{FIG-toric}. Therefore, we conclude that to observe the
toric code on small lattices, one has to limit residual
interactions up to a few percent of the coupling strength $J$.

This work is supported by Microsoft-Q, DARPA-QUEST, NSF-PFC-JQI,
and ARO-DARPA.
%
%


\end{document}